\documentclass[aps,prb,twocolumn,showpacs,preprintnumbers,amsmath,amssymb,superscriptaddress]{revtex4-1}
\usepackage{graphicx}
\usepackage{dcolumn}
\usepackage{bm}
\usepackage{color}
\usepackage{ulem}

\newcommand{\bra}[1]{\langle #1|}
\newcommand{\ket}[1]{|#1\rangle}

\newcommand{\expect}[1]{\langle #1 \rangle}

\newcommand{\be}{\begin{equation}}
\newcommand{\ee}{\end{equation}}
\newcommand{\bea}{\begin{eqnarray}}
\newcommand{\eea}{\end{eqnarray}}
\newcommand{\eq}[1]{Eq.~(\ref{#1})}
\newcommand{\fig}[1]{Fig.~\ref{#1}}

\newcommand{\e}{\varepsilon}

\newcommand{\s}{\sigma}
\newcommand{\G}{\Gamma}

\newcommand{\tSO}{t_{\rm SO } }

\begin{document}

\title{Transport through graphene-like flakes with intrinsic spin orbit interactions}

\author{I. Weymann}
\email{weymann@amu.edu.pl} \affiliation{Faculty of Physics,
Adam Mickiewicz University, 61-614 Pozna\'n, Poland}

\author{J. Barna\'s}
\affiliation{Faculty of Physics, Adam Mickiewicz University,
61-614 Pozna\'n, Poland} \affiliation{Institute of Molecular
Physics, Polish Academy of Sciences, 60-179 Pozna\'n, Poland}

\author{S. Krompiewski}
\affiliation{Institute of Molecular Physics, Polish Academy of
Sciences, 60-179 Pozna\'n, Poland}

\date{\today}

\begin{abstract}
It has been shown recently [J. L. Lado et al., Phys. Rev. Lett. {\bf 113},
027203 (2014)] that edge magnetic moments in graphene-like
nanoribbons are strongly influenced by the intrinsic spin-orbit
interaction. Due to this interaction an anisotropy comes about
which makes the in-plane arrangement of magnetic moments
energetically more favorable than that corresponding to the
out-of-plane configuration. In this paper we raise both the edge
magnetism problem as well as differential conductance and shot
noise Fano factor issues, in the context of finite-size
flakes within the Coulomb blockade (CB) transport regime. Our
findings elucidate the following problems: (i) modification of the
CB diamonds by the appearance of the in-plane magnetic moments,
(ii) modification of the CB diamonds by intrinsic spin-orbit interaction.
\end{abstract}




\maketitle

\section{Introduction}

Graphene-like nanostructures -- including silicene
\cite{Guzman2007}, germanene \cite{Cahangirov2009} and
hypothetically also stanene (two-dimensional tin) \cite{Liu2011}, aluminene
\cite{Kamal2015} and others -- constitute a particularly
challenging class of materials potentially important for future
applications in nanoelectronics and spintronics.
On the one hand,
the great interest in these materials results from many
fascinating properties of graphene \cite{Geim2007}, which have
been discovered and reported in the last decade. It is believed
that these new nanomaterials akin to graphene as regards the
honeycomb atomic structure and possibly also the Dirac fermion
features, will also share well-known graphene's superlatives
concerning mechanical, optical, electrical, magnetic and thermal
properties. \cite{NetoRMP09}
Due to the nonequivalence of the involved
sublattices (buckling) and the possibility to engineer an energy
band gap, \cite{Zandvliet2014}  the new nanostructures, in contrast
to graphene, may prove to be useful for realization of the field
effect transistor. \cite{Tao15}

Some other interesting properties of these two-dimensional materials follow from internal spin-orbit interaction. \cite{Ezawa}
In the bulk (two-dimensional) limit, this interaction opens a gap at the Fermi level
in the electronic spectrum of the corresponding Dirac fermions.
This gap leads to the spin Hall insulator phase.\cite{Kane2005}
Moreover, the induced gap can be tuned by an external electric field
oriented perpendicularly to the system's plane.\cite{Dyrdal,Drummond12}
Unfortunately, the intrinsic spin-orbit interaction is rather
small in these materials, therefore the gap is hardly resolved in experiments.
Additionally, the spin-orbit coupling can give rise to an in-plane magnetic anisotropy,
so that the edge magnetic moments induced by Coulomb
interactions become oriented in the layer plane.\cite{Lado2014}

The electronic spectrum, and especially the energy gaps in the spectrum,
can also be controlled by other means.
It is well known that a key factor responsible for opening  energy gaps is
the reduced geometry (confinement effect). This, in turn, has a significant influence on
transport properties, as shown for instance in the case of long silicene nanoribbons~\cite{Lou2009,Zhang2010,Lopez2013,Zberecki} and graphene quantum dots.\cite{Weymann2012}
Remarkable impact on the electronic structure is also due to the edges, e.g. zigzag \textit{vs.}
armchair ones. Moreover, electron correlations, especially of Hubbard-type,
also play a role in the formation of spin-dependent electronic states.

In this paper we study the effects of spin-orbit interaction
on electronic and transport properties of small graphene-like
flakes, in which size quantization in both directions occurs.
The corresponding electronic spectrum of such
{\it graphene-like quantum dots} (GLQDs) is then discrete.
To model the system, apart from the on-site Coulomb interaction (of Hubbard type),
one then also has to account for the Coulomb blockade effects,
especially when the coupling of the dot to external leads is relatively weak, as considered in this work.
With the aid of the exact diagonalization, we first find the eigenstates and eigenvalues
of isolated GLQD, which are then used
to determine the transport characteristics by using
the real-time diagrammatic technique in the lowest order
perturbation expansion with respect to the coupling to external leads.
\cite{diagrams,thielmann,weymannPRB08}
In particular, we calculate the differential conductance as well as the shot noise
in both the linear and nonlinear response regimes.
Since the conductance and Fano factor provide information
about the electronic spectrum of the system,
and the electronic states are modified by the Coulomb and spin-orbit interactions,
the calculated transport characteristics {\it vs} gate and transport voltages
allow us to draw some conclusions on these interactions.

The format of the paper is as follows.
In Sec. \ref{Sec:GLNRprop} the Hamiltonian (Sec. \ref{Sec:H})
and electronic structure (Sec. \ref{Sec:spectra}) of GLQD are presented.
The importance of the intrinsic spin-orbit interaction and its
effect on the appearance of the in-plane anisotropy is emphasized,
and the results on edge magnetism and energy
spectra of an isolated GLQD are presented.
Transport properties are considered in Sec. \ref{Sec:transport}.
The effective Hamiltonian and method for studying transport properties
are described in Secs. \ref{Sec:Heff} and \ref{Sec:RTDT}, respectively,
while transport characteristics are analyzed in Sec. \ref{Sec:results}.
Finally, Sec. \ref{Sec:sum} summarizes the main findings.

\section{Electronic properties}\label{Sec:GLNRprop}

In the following we analyze the electronic properties
of an isolated quantum dot of graphene-like materials.
We first present the Hamiltonian of GLQD in the mean field approximation.
Then, by means of exact diagonalization,
we find the eigenstates and eigenvalues of the dot
and analyze its electronic and spin properties.

\subsection{Hamiltonian}\label{Sec:H}

The system under consideration is described by a tight-binding Hamiltonian which also contains,
apart from the usual hopping term, the intrinsic spin-orbit
and the Hubbard correlation terms.
The latter is treated within the mean-field approximation.
Thus, the Hamiltonian can be written as
\be \label{H}
H = H_{\rm hop} + H_{\rm U},
\ee
where the first term is given by
\begin{eqnarray} \label{Ht}
  H_{\rm hop} \!= \! - \!\!\!\! \sum \limits_{ <i j>,\sigma } \!\!\! t_{ij} d_{i\sigma}^\dagger d_{j\sigma}
  \!+\! i \, \tSO \!\!\!\! \sum \limits_{ <\!\!<ij>\!\!>} \!\!\!
  \nu_{ij} ( d_{i\uparrow}^\dagger d_{j\uparrow}
  \! -\! d_{i\downarrow}^\dagger d_{j\downarrow}),
\end{eqnarray}
while the second term takes into account
the Coulomb correlations in the mean-field approximation
and has the form
 \begin{eqnarray} \label{HU}
H_{\rm U}&=& U \sum \limits_{i} \left(
\expect{n_{i \downarrow}} n_{i\uparrow}
+ \expect{n_{i \uparrow}} n_{i \downarrow}
- \expect{n_{i \uparrow}} \expect{n_{i\downarrow}} \right. \nonumber \\
&&\left. - \expect{S_i^+} S_i^- - \expect{ S_i^-} S_i^ + + \expect{S_i^+ } \expect{ S_i^-} \right).
\end{eqnarray}
Here, $n_{i \sigma}$ are the corresponding occupation operators,
$n_{i \sigma} = d_{i\sigma}^\dagger d_{i\sigma}$, while
$S_i^+ = d_{i \uparrow}^\dagger d_{i\downarrow}$ and
$S_i^- = d_{i\downarrow}^\dagger d_{i\uparrow}$,
with $d_{i\sigma}^\dagger $ ($ d_{i\sigma} $)
being the creation (annihilation) operator of spin-$\sigma$ $\pi$-electrons at a lattice point $i$.
The angle brackets stand here for the expectation values calculated
with respect to the ground state of the Hamiltonian $H$.
Moreover, $U$ is the Hubbard parameter of the on-site repulsion, $\tSO$ denotes the
intrinsic spin-orbit parameter, whereas $t_{ij}$ and $\nu_{ij}$
are the hopping integrals and the Haldane factors between nearest
neighbor and next nearest neighbor sites, respectively.
For a pair of next nearest sites, $i$ and $j$, with a common nearest neighbor $k$,
$\nu_{ij}=(\vec{r}_{ik} \times \vec{r}_{jk}) / \vert
\vec{r}_{ik} \times \vec{r}_{jk} \vert$. \cite{HaldanePRL88,Kane2005}
For simplicity, the hopping integrals $t_{ij}$ are assumed to be
nonzero only for nearest neighbors, and the corresponding
hopping parameter $t$ is used as the energy unit (e.g. $t=$ 2.7, 1.5,
1.4, 1.3 eV for graphene, silicene, germanene and stanene
[\onlinecite{Lado2014}], respectively).

It is noteworthy that the correlation part of the Hamiltonian,
$H_{\rm U}$, comes from the full Hartree-Fock decoupling
of the Hubbard Hamiltonian. Usually only
the first spin-diagonal part of \eq{HU} is taken into account.
\cite{Fernandez07,Weymann2012, SK1214, Szal15}
However, in the present anisotropic case the magnetization direction may be
arbitrary, so the spin mixing term must not be skipped.
Similar Hamiltonians, in other contexts, were studied \textit{i.a.} in Refs.
[\onlinecite{Rhim2009, Lado2014a}].

In numerical calculations the Coulomb on-site repulsion is assumed to be
$U = t$ (cf. Ref. [\onlinecite{Lado2014}]).
The expectation values of the relevant occupation numbers, $\expect{n_{i \sigma}}$,
have been computed self-consistently by summing up the squared
eigenvectors corresponding to the eigenvalues not greater than the Fermi energy.
Furthermore, for the out-of-plane and in-plane configurations
the following order parameters come into play:
$M_z^i = \mu_B \expect{n_{i \uparrow}-n_{i\downarrow}}$ and $M_x^i=\mu_B
\expect{S_i^+ + S_i^-}$, where  $\mu_B$ is the Bohr magneton.

\subsection{Electronic and spin structure}\label{Sec:spectra}

\begin{figure}[t!]
  \includegraphics[width=0.8\columnwidth]{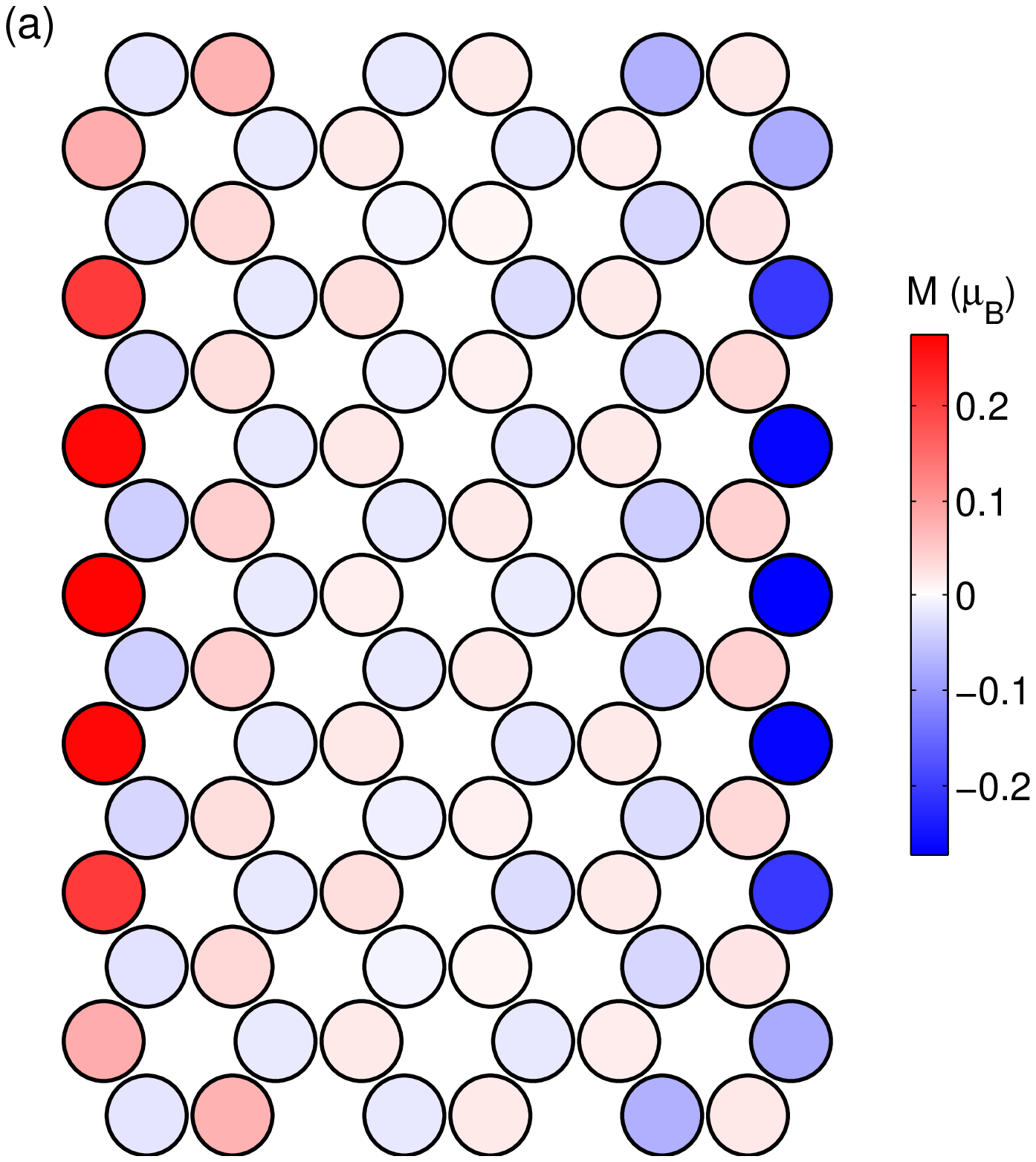}
  \includegraphics[width=0.8\columnwidth]{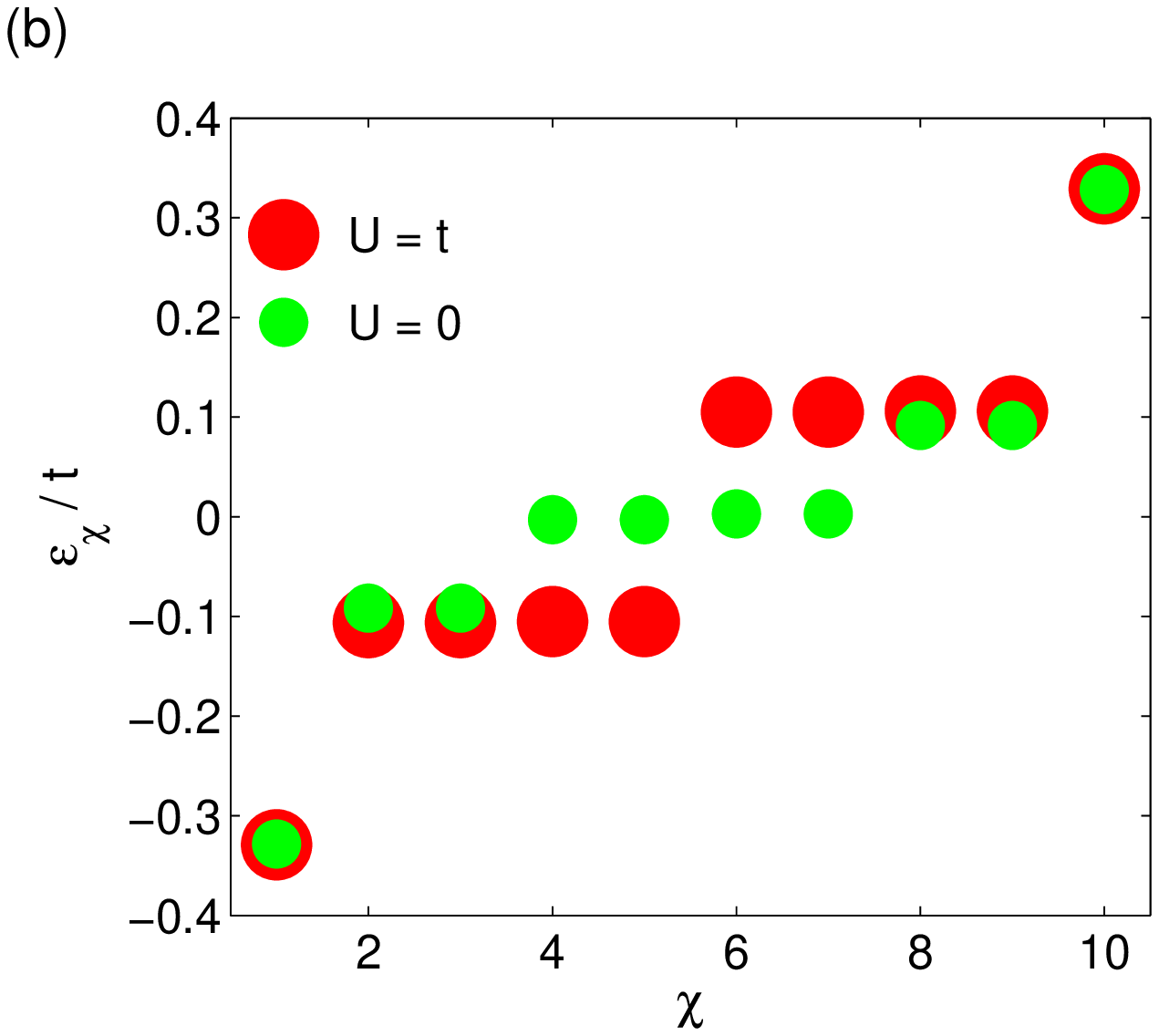}
  \caption{\label{Fig:2}
  (Color online)
  Distribution of magnetic moments on individual atoms (a),
  and the eigenvalues (b) around the charge neutrality point
  calculated for the case of $U=t$ and $\tSO=0$.
  Because $\tSO=0$, the GLQD magnetization is isotropic.
  For comparison, in panel (b) we also show the eigenvalues calculated for $U=0$.
  In this case the magnetic moments do not form
  and there is no energy gap either.}
\end{figure}

To exemplify our results we consider a relatively small
zigzag-type rectangular GLQD containing $90$ atoms, see \fig{Fig:2}(a).
The model system is a honeycomb structure as that of
graphene, but in contrast to graphene, now the sublattices $A$ and $B$
are vertically shifted with respect to each other by the distance
of the order of 0.5 {\AA}. \cite{Cahangirov2009}
Incidentally, we have checked by performing additional computations
for some longer systems, that this model is large enough
to yield correct magnetic moments at its mid-length.
The buckling of the considered graphene-like flake
is closely related with the intrinsic spin orbit
interaction -- if buckling is negligible, then $\tSO\approx 0$,
and one obtains the case of graphene.

Since the electronic and magnetic properties of the considered GLQD
depend greatly on both the on-site Coulomb correlations
and spin-orbit interactions, to elucidate the role of those interactions,
in the following we present and discuss
results for the case of $U=0$ and $U>0$, as well as
$\tSO = 0$ and $\tSO>0$.
As shown below, finite on-site Coulomb correlations
are responsible for magnetic moments at individual atoms
that form at the edges (edge magnetism). On the other hand,
the spin-orbit interaction determines
the magnetization configuration and its anisotropy.

\subsubsection{Absence of spin-orbit interaction}

The distribution of magnetic moments at individual
atoms of the considered flake in the absence
of spin-orbit interaction and for $U=t$ is shown in \fig{Fig:2}(a).
In the case of $\tSO = 0$, the magnetization is isotropic.
It is readily seen that magnetic moments have
relatively high values (roughly 0.3 $\mu_{B}$) at the zigzag
edges, except at atoms located close to the armchair edges.
Moreover, while atoms belonging to one sublattice are magnetized in one direction,
the magnetization of atoms from the other sublattice is opposite, see \fig{Fig:2}(a).
This antiparallel magnetic configuration has got a lower
energy than the parallel one.

The eigenenergies of the GLQD Hamiltonian are shown
in \fig{Fig:2}(b) in the case of both finite $U$ and $U=0$.
This figure presents discrete energy levels
$\varepsilon_\chi$ around the Fermi level
(defined by the charge neutrality point, $E_F = 0$),
with $\chi$ ranging from 1 to 10.
These energy levels will be later used for studying the transport properties.
Due to the Hubbard correlations, an energy gap develops,
see the large points in \fig{Fig:2}(b). However, if the
correlations disappear, $U=0$ [see the small points in \fig{Fig:2}(b)],
the energy gap closes at the charge neutrality point, and degenerate energy
states appear at the Fermi energy.

\subsubsection{Finite spin-orbit interaction}

\begin{figure}[t]
  \includegraphics[width=0.8\columnwidth]{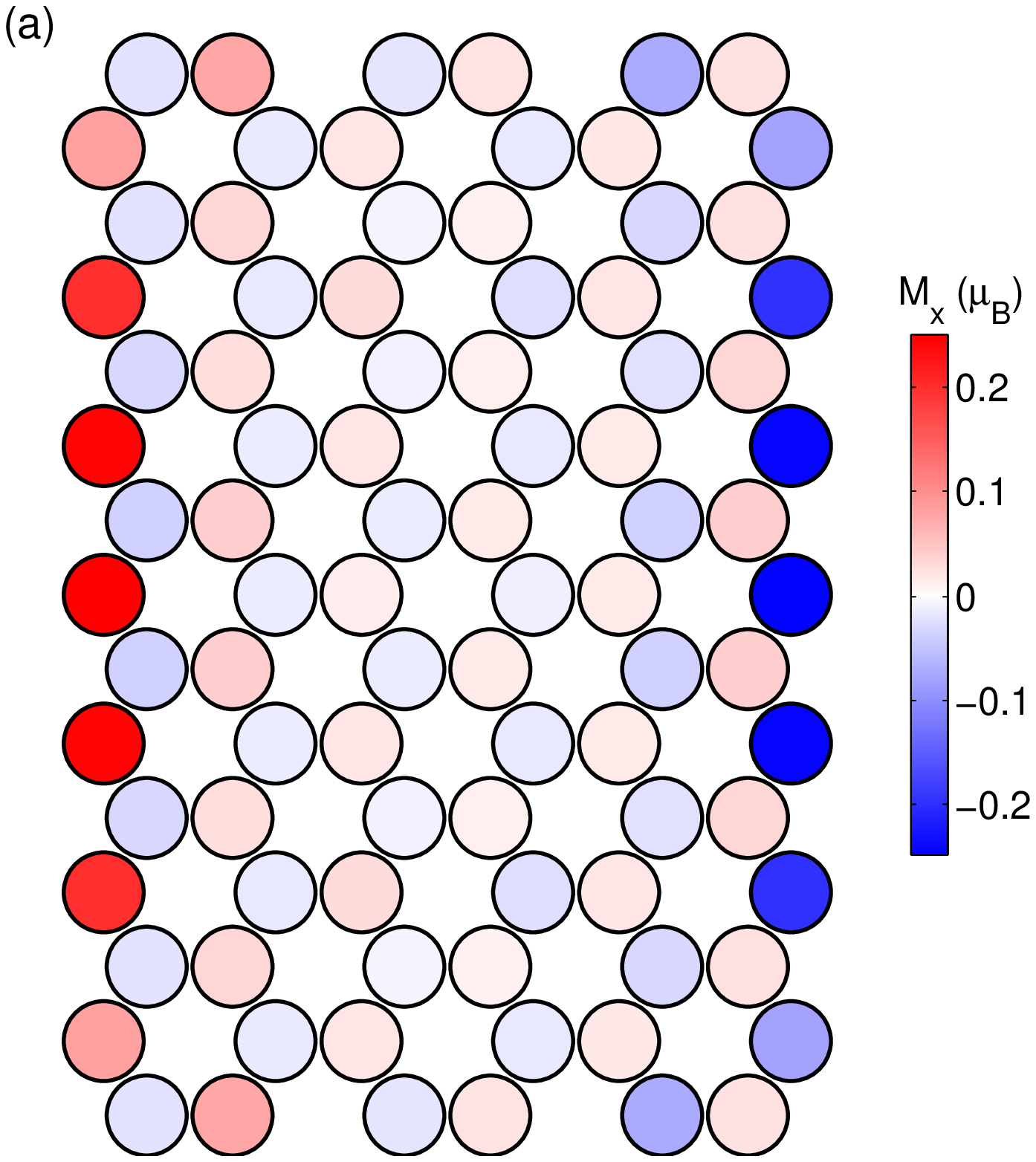}
  \includegraphics[width=0.8\columnwidth]{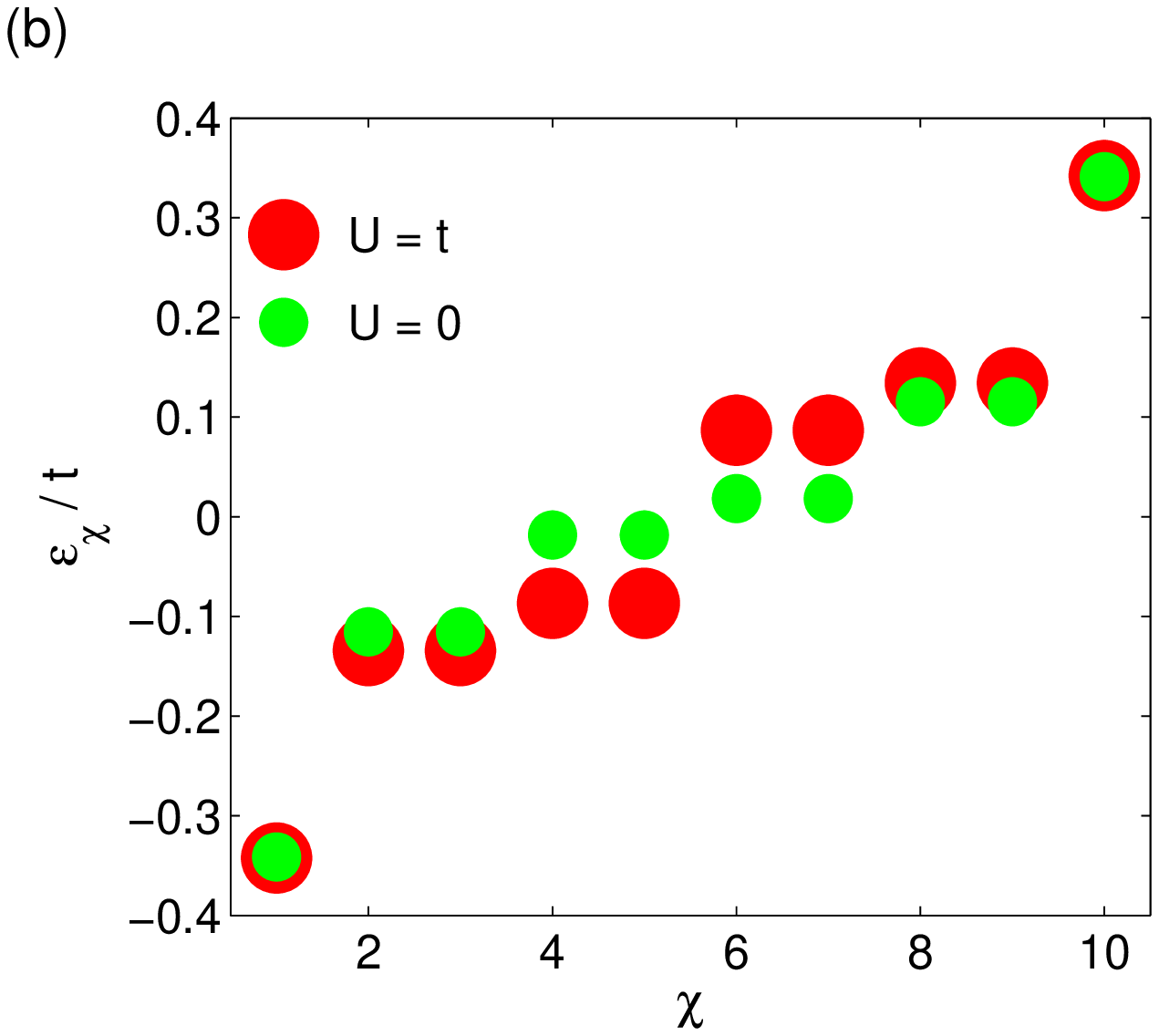}
  \caption{\label{Fig:3}
  (Color online) The same as in \fig{Fig:2} calculated for
  finite spin orbit interaction, $\tSO/t=0.025$.
  The ground state magnetic configuration
  corresponds to the in-plane arrangement,
  and the mid-length edge magnetic moments
  are reduced by more than 9\% with respect
  to those in the case of $\tSO=0$ shown in \fig{Fig:2}(a).}
\end{figure}

The situation changes when the spin-orbit interactions are present,
since now magnetic anisotropy develops.
As a result, magnetic configurations corresponding to
the in-plane and the out-of-plane orientations differ from each other.
The calculations show that energetically more favorable
is the configuration with the in-plane easy axis.
The obtained magnetization profile is shown \fig{Fig:3}(a).
In comparison with \fig{Fig:2}(a), the magnitude of magnetic
moments is now slightly diminished.
Moreover, similarly as in the case of $\tSO = 0$,
the direction of magnetic moments is opposite
for atoms belonging to different sublattices.
The corresponding eigenenergies around the Fermi level are shown in \fig{Fig:3}(b).
One can see that now there appears a small energy gap of purely
spin-orbit origin when $U=0$ [the small points in \fig{Fig:3}(b)].
The energy gap gets pronouncedly increased due to an extra magnetic
contribution for $U>0$, see the large points in \fig{Fig:3}(b).

Figures \ref{Fig:2} and \ref{Fig:3} show that the dominant
contribution to the band gap is magnetic in nature
and results from finite Coulomb correlations.
The intrinsic spin-orbit contribution is competitive
with the magnetic one and, consequently, the net energy gap
gets reduced while increasing the spin-orbit parameter $\tSO$,
cf. Figs. \ref{Fig:2}(b) and \ref{Fig:3}(b).
Interestingly, a similar tendency can also be deduced
from the results presented in Ref. [\onlinecite{EzawaPRL12}]
for the out-of-plane configuration.
Moreover, the aforementioned behavior resembles to some extent
the competition between a vertical electric field and the
intrinsic spin-orbit interaction, as reported recently.
\cite{Drummond12,EzawaPRL12,DrummondPRB13}

\begin{figure}[t]
  \includegraphics[width=0.9\columnwidth]{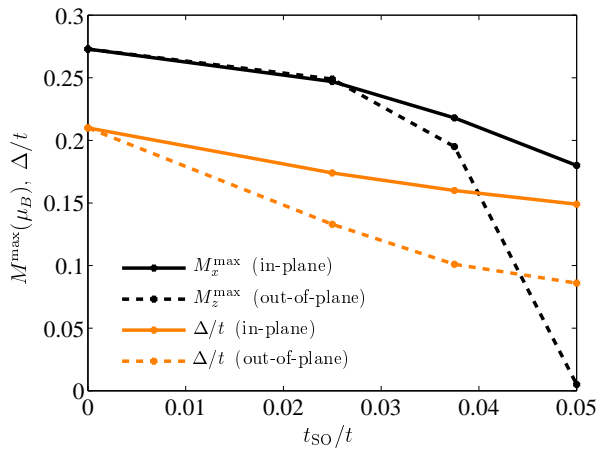}
  \caption{\label{Fig:4}
  (Color online) Maximum edge magnetic moments $M^{\rm max}$
  and the corresponding energy gaps $\Delta$
  plotted as a function of the intrinsic spin-orbit coupling parameter $\tSO$ for $U=t$.
  The stable (in-plane configuration) data are plotted with solid lines.
  For comparison, the unstable out-of-plane data are also shown (dashed lines).}
\end{figure}

A more detailed analysis of the effect of the intrinsic spin-orbit coupling on the maximum
(mid-length) magnetic moment ($M^{\rm max}$) and the corresponding energy gap
($\Delta$) is depicted in \fig{Fig:4}. The data were calculated for $U=t$
and are shown for both in-plane and out-of-plane configurations.
As can be seen in the figure, both $M^{\rm max}$ and $\Delta$ decrease with
increasing the spin-orbit coupling parameter $\tSO$.
As mentioned above, this decrease is related with
competition between spin-orbit and magnetic contributions.
For larger values of the spin-orbit coupling, these two
quantities are even more strongly affected in the
unstable out-of-plane configuration, see \fig{Fig:4}.

We would like to note that for the considered rectangular graphene-like flakes with
both zigzag and armchair edges, our numerical results show that noticeable magnetic moments
can only appear on edges of the former type.
Moreover, due to the spin-orbit interaction these 
moments are oriented in the plane of the flakes.
The question which now arises is whether these observations
are more general and also hold for other shapes of the flakes. 
This problem in the absence of spin-orbit interaction
was considered e.g. in Refs. [\onlinecite{Weymann2012,Kikutake,Rakyta}], 
from which follows that localized edge magnetic moments
exist mainly on zigzag-like fragments of an arbitrary boundary.
To address this issue in the presence of spin orbit interaction
in the Appendix we consider a flake of circular shape exhibiting some irregularities.
Our results suggest that the above observation is
indeed applicable for more general shapes of graphene-like flakes.
Moreover, we also show that the magnetic moments
are oriented in the plane of the flake,
similarly to the case of rectangular flakes.

\section{Transport properties}\label{Sec:transport}

In this section we focus on the analysis of transport properties
of GLQD attached to external electrodes.
We assume that the coupling between the dot and external
leads is relatively weak, so transport occurs mainly due to
sequential tunneling processes. To determine transport characteristics,
we employ the real-time diagrammatic technique in the lowest order
of perturbation expansion with respect to the coupling strength between GLQD and the leads.

\subsection{Effective Hamiltonian}\label{Sec:Heff}

From exact diagonalization of the GLQD's Hamiltonian (\ref{H}) we obtain the
eigenvalues $\varepsilon_{\chi}$ and eigenstates
$\ket{\Psi_{\chi}} = \sum_{i\s}u_{i\s}\ket{i\s}$. Since in our
analysis we are interested in the low bias voltage regime, in
calculations we will consider only limited number of eigenstates
with eigenvalues close to the charge neutrality point. The
Hamiltonian of the isolated GLQD can be written as
\begin{equation} \label{HGQD}
  H_{\rm GLQD} = \sum_{\chi} \varepsilon_{\chi} d_\chi^\dag d_\chi
  + \frac{E_C}{2} \left(N - N_0 \right)^2 \,,
\end{equation}
where $d_\chi^\dag$ is the creation operator for an electron of energy $\e_\chi$.
Note that in general the states $\ket{\chi}$ are linear combinations
of both spin-up and spin-down states when the quantization axis is normal to the system's plane.
The charging energy of the graphene-like quantum dot is denoted by $E_C$,
$N = \sum_\chi d_\chi^\dag d_\chi$,
while $N_0$ denotes the number of electrons
in the electrically neutral dot.

The Hamiltonian of electrodes, in turn, takes the form
\begin{equation}
  H_{\rm Leads} = \sum_{r=B,T}
  \sum_{{\bf k},\sigma} \varepsilon_{r {\bf k} \sigma} c^\dag_{r {\bf k}
  \sigma} c_{r {\bf k} \sigma}
\end{equation}
and models the leads as reservoirs of noninteracting
quasi-particles. Here, $c^\dag_{r {\bf k}\sigma}$ is the creation
operator for a spin-$\sigma$ electron with wave vector ${\bf k}$
in the bottom ($r=B$) or top ($r=T$) lead, while $\varepsilon_{r {\bf k}\sigma}$ denotes the
corresponding energy.

The tunneling processes between the GLQD and the leads
can be described by the following tunneling Hamiltonian
\begin{equation}
  H_{\rm Tun} = \sum_{r {\bf k} \sigma } \sum_{\chi} \sum_\eta
  v_{r}^{(\eta)} \alpha_{r\chi \sigma}^{(\eta)}
  \left[
     c_{r {\bf k} \sigma}^\dag  d_\chi +  d_\chi^\dag c_{r {\bf k} \sigma}
  \right],
\end{equation}
where $v_{r}^{(\eta)}$ is the hopping matrix element between
the dot and the lead $r$, which is assumed to be momentum and spin independent.
The coefficient $\alpha_{r\chi \sigma}^{(\eta)}$ is defined as
\be
  \alpha_{r\chi \sigma}^{(\eta)} = \sqrt{{\sum_{i}}^{(\eta)} |u_{i\s}|^2}\; ,
\ee
where $(\eta)$ means that the summation is over the $\eta$-th row of atoms
counted from bottom ($r=B$) or top $(r=T)$ side of the flake,
which is attached to the electrodes.
If one includes all the atoms, then $\alpha_{r\chi \sigma}^{(\rm all)} = 1$.
The effective broadening of the  GLQD's levels
can be described by
\be
   \G_{r\chi} = \sum_{\s\eta} 2\pi \rho_{r} |\alpha_{r\chi \sigma}^{(\eta)}|^2 |v_r^{(\eta)}|^2 =
   \sum_{\s\eta} |\alpha_{r\chi \sigma}^{(\eta)}|^2 \Gamma_{r}^{(\eta)} ,
\ee
with $\rho_{r}$ being the density of states of lead $r$ and
$\Gamma_{r}^{(\eta)} = 2\pi \rho_{r}  |v_r^{(\eta)}|^2$.

In calculations we consider the coupling to the first
row of atoms next to the contacts, that is $\eta = 1$, and assume that the tunnel
matrix elements $v^{(\eta)}_r$ for $\eta>1$ are negligible.
This assumption is justifiable since
tunnel matrix elements depend exponentially on the distance,
and the coupling to next rows of atoms is expected to change the GLQD level widths
only insignificantly.
We thus assume ($\eta=1$): $\Gamma_{B}^{(\eta)} =
\Gamma_{T}^{(\eta)} \equiv \Gamma/2$.
The charging energy of the graphene-like quantum dot is estimated
to be  $E_C/t = 0.1$ (see Ref. [\onlinecite{Ma13}] for details concerning graphene QD).
Moreover, in calculations we restrict ourselves to the low energy regime
and take into account $8$ states of GLQD around the Fermi level.
\cite{Weymann2012}

\subsection{Method}\label{Sec:RTDT}

To determine the transport characteristics we make use of
the diagrammatic technique in real time.
\cite{diagrams,thielmann}
Within this approach, one performs a systematic perturbation expansion
of the reduced density matrix and relevant operators
with respect to the tunneling processes between the nanostructure and the leads.
In the following analysis we focus on the weak coupling regime
and assume that the main contribution to the conductance
is captured by the lowest-order tunneling processes,
where electrons tunnel sequentially, one by one, through the junction.
In the considered case the density matrix is diagonal
and its elements, which directly correspond to the occupation probabilities $P_\chi$
of respective GLQD states $\ket{\chi}$,
can be found from appropriate kinetic equation. \cite{diagrams}
In the steady state, one has \cite{diagrams,thielmann}
\be
  \sum_{\chi\chi'} W_{\chi\chi'} P_{\chi'} = 0,
\ee
together with normalization condition, $\sum_\chi P_\chi = 1$.
Here, ${\mathbf W}$ is the self-energy matrix whose elements,
$W_{\chi\chi'}$, describe transitions between respective states
$\ket{\chi}$ and $\ket{\chi'}$. These matrix elements
can be found using the respective diagrammatic rules
\cite{diagrams,thielmann,weymannPRB08}
and are given by
$W_{\chi\chi'} = W_{\chi\chi'}^B + W_{\chi\chi'}^T$, with
\begin{eqnarray*}
   W_{\chi \chi^\prime}^r &=& 2\pi \sum_{\sigma} \rho_r
   \Big\{
   f_r(\varepsilon_\chi - \varepsilon_{\chi^\prime})
   \Big| \sum_{\eta \chi''}
   v_{r}^{(\eta)} \alpha_{r\chi\sigma}^{(\eta)} \bra{\chi}d_{\chi''}^\dagger
   \ket{\chi^\prime}\Big|^2 \\
   &+& \left[ 1-f_r(\varepsilon_{\chi^\prime} - \varepsilon_{\chi}) \right]
   \Big|\sum_{\eta \chi''} v_{r}^{(\eta)} \alpha_{r\chi\sigma}^{(\eta)} \bra{\chi}d_{\chi''}
   \ket{\chi^\prime}\Big|^2 \Big\},
\end{eqnarray*}
for $\chi \neq\chi'$ and $W_{\chi\chi}^r = -\sum_{\chi^\prime \neq \chi} W_{\chi^\prime\chi}^r$.
Here, $f_r(\e) = f(\e-\mu_r)$ is the Fermi-Dirac distribution
function and $\mu_r$ denotes the electrochemical potential of lead $r$.
The current flowing through the system can be found from
\cite{diagrams,thielmann}
\be
  I = \frac{e}{2 \hbar} \sum_{\chi\chi'} W^I_{\chi\chi'} P_{\chi'}
\ee
where the elements $W^I_{\chi\chi'}$ are essentially given by
$W_{\chi\chi'}$ but this time they take into account
the number of electrons transferred through the system.
\cite{diagrams,thielmann}
In addition, to gain more detailed understanding of transport properties,
we also study the behavior of the Fano factor,
$F = S / S_P$, which measures the deviation of the
zero-frequency shot noise $S$ from the Poissonian
shot noise given by, $S_P = 2|eI|$.
\cite{blanterPR00}
The formula for the calculation of shot noise using
the diagrammatic technique can be found in Ref. [\onlinecite{thielmann}].

\subsection{Numerical results}\label{Sec:results}

Using the above described formalism we have computed the differential
conductance density plots, i.e. the differential conductance
$G=dI/dV$ vs. bias voltage $V$ and gate voltage
(effectively the position of the electronic spectrum of GLQD denoted by $\e$),
as well as the density plots of the corresponding shot noise Fano factor.
In the following we focus mainly on analyzing the effects of the on-site Coulomb interaction
(edge magnetism) and spin-orbit coupling on transport characteristics.
To analyze the role of edge magnetism, we consider
first the situation with zero spin-orbit coupling, $\tSO=0$.

\subsubsection{No spin-orbit interaction}

\begin{figure}[t!]
  \includegraphics[width=1\columnwidth]{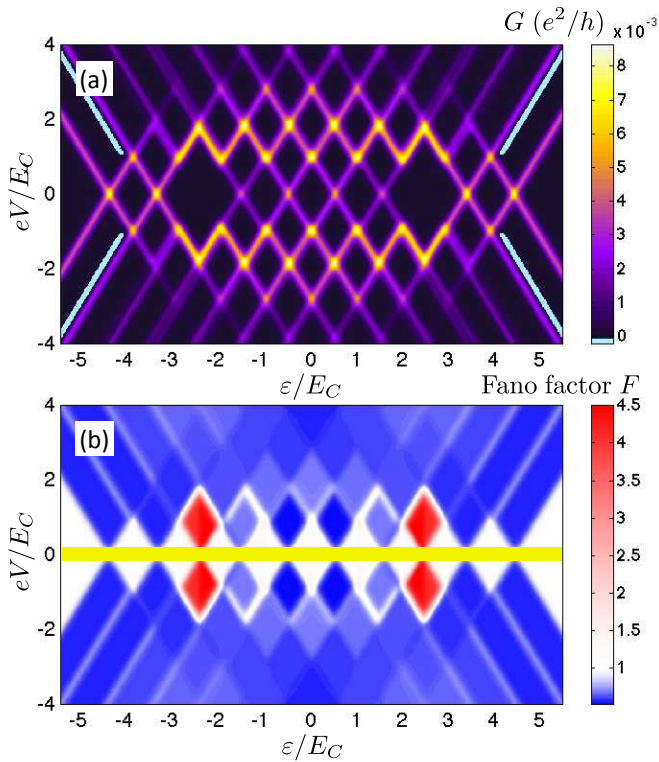}
  \caption{\label{Fig:5}
  (Color online) The differential conductance (a)
  and the shot noise Fano factor (b) as a function
  of the bias voltage $V$ and the GLQD level position $\e$
  calculated for nonmagnetic GLQDs ($U=0$) and
  in the absence of spin-order coupling ($\tSO=0$).
  The Fano factor is plotted in appropriately tuned scale,
  such that white color corresponds to $F=1$.
  Since the Fano factor is divergent in the low bias voltage regime,
  this transport region is covered by a yellow stripe.
  Negative differential conductance is marked with cyan (bright) color.
  The parameters are:
  $E_C/t=0.1$, $\Gamma /E_C = 0.01$
  and thermal energy $k_B T/E_C=0.05$.
  }
\end{figure}

\begin{figure}[t!]
  \includegraphics[width=1\columnwidth]{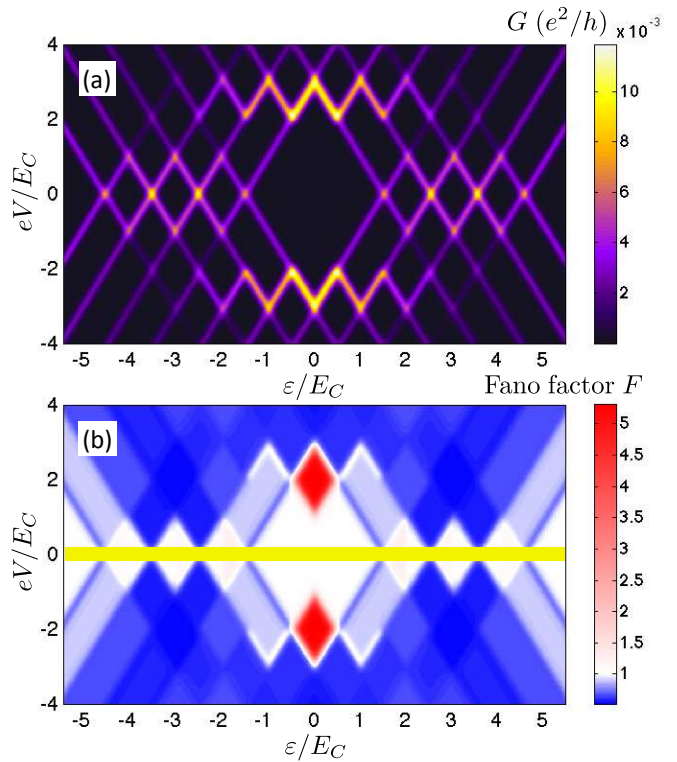}
  \caption{\label{Fig:6}
  (Color online) The differential conductance (a)
  and shot noise Fano factor (b) spectra
  in the case of magnetic, $U=t$, GLQDs and for
  vanishing spin-orbit interaction, $\tSO=0$.
  The other parameters are the same as in \fig{Fig:5}.
}
\end{figure}

The numerical results on the differential conductance and Fano factor
for $U=0$ and $U=t$ are presented in Figs. \ref{Fig:5} and \ref{Fig:6}, respectively.
In both cases the spin-orbit interaction was assumed to be equal to zero, $\tSO=0$.
Since the Coulomb interaction leads to edge magnetic moments,
the latter figure also illustrates effectively the impact of
the edge magnetism on both the Coulomb blockade (conductance)
and the Fano factor spectra.

The bias and gate voltage dependence of the differential conductance
in the case of $U=0$ is presented in \fig{Fig:5}(a).
This figure shows the stability diagram of the device, with
Coulomb diamonds visible at low bias voltage.
In diamonds the current through the system is suppressed by the charging energy
and only thermally-activated or higher-order tunneling events are possible.
The rate of sequential processes increases
once the voltage approaches a threshold voltage,
leading to a step in the current and associated peak in the differential conductance.
Next steps occur with further increasing the bias voltage,
leading to additional lines in $G$, see \fig{Fig:5}(a).
Since from the size of Coulomb diamonds one can extract information
on the relevant energy scales, let us discuss it in more detail.

In the case of $U=0$, there are four roughly degenerate
states of $\varepsilon \approx 0$ in an isolated dot, see \fig{Fig:2}.
The energy difference between these dot levels
is resolved neither in \fig{Fig:2}(b) nor in \fig{Fig:5}(a).
These four discrete states are responsible for the three central diamonds visible in \fig{Fig:5}(a),
which are equal in size. The middle diamond refers to the charge neutrality point,
when the two discrete zero-energy levels out of the four ones are occupied by electrons
and the dot is then electrically neutral.
Considering $\varepsilon$ in \fig{Fig:5} (and also in other figures) as an effective gate voltage,
the next discrete state becomes occupied when the energy shift
due to the gate voltage is equal to $\e = E_C/2$.
The dot becomes then charged with one excess electron.
In turn, the fourth discrete level becomes populated when
the energy shift due to the gate voltage is equal to $\e = (3/2)E_C$,
so that two excess electrons can reside on the dot for $(3/2)E_C < \varepsilon < (5/2)E_C$.
When the gate voltage increases further,
then the third excess electron can enter the dot,
and this electron will occupy one of the two levels of higher (positive) energy [cf. \fig{Fig:2}(b)].
The additional shift due to the gate voltage must be now larger,
as it is equal to the charging energy $E_C$ plus the corresponding level spacing.
This corresponds to the first large diamond on the positive energy side of \fig{Fig:5}(a).
The next diamond is of the same size as the central one
since the corresponding gate shift is again equal to $E_C$.
Similar scenario also holds when the gate voltage changes sign,
except that now the number of excess electrons becomes negative.
Since the energy spectrum is symmetrical around $\e = 0$ due to the particle-hole symmetry,
the whole spectrum in \fig{Fig:5}(a) is also symmetrical
with respect to the sign change of $\varepsilon$.

The corresponding Fano factor is shown in \fig{Fig:5}(b).
When $F=1$, the shot noise is Poissonian,
while for $F>1$ ($F<1$) the noise is super-Poissonian (sub-Poissonian).
To facilitate identification of different behavior of the Fano factor,
figures presenting density plots of $F$ are plotted in appropriately tuned scale.
Moreover, because the Fano factor is divergent when $V\to 0$,
this transport region is covered by a yellow stripe.
As can be seen in \fig{Fig:5}(b),
the shot noise is predominantly super-Poissonian, $F>1$,
in the Coulomb blockade regions (diamonds),
where the sequential transport is suppressed by the Coulomb interaction.
Only in small parts of the diamonds the noise is close to Poissonian.
On the other hand, beyond the Coulomb blockade regions,
the shot noise is sub-Poissonian, $F<1$,
which is typical for Coulomb-correlated transport.
We note that transport in the Coulomb blockade regime may not be described quite well
by the first-order (sequential) processes, and to describe it more accurately,
one would need to go beyond the sequential tunneling approximation
and include also cotunneling processes. \cite{cotunneling}
Nevertheless, since in our considerations we assume a very weak coupling to external leads,
our results can still be considered as qualitatively sound.

We note that there is another interesting feature visible in \fig{Fig:5},
which is associated with negative values of the differential conductance
that develop in some transport regions.
In fact, the most pronounced negative differential conductance can be seen
when $\e/E_C \gtrsim |4|$, emphasized in \fig{Fig:5}(a) with the cyan (bright) stripes.
This effect occurs when discrete energy levels of GLQD
are coupled to the leads with considerably different strength
and the difference between the couplings occurs for consecutive levels,
for which the level spacing is much larger than thermal energy.
This happens in the case when $\tSO = U = 0$,
therefore only in this situation we observe the effect
of negative differential conductance.
More specifically, for e.g. $\e/E_C \approx |5|$,
with increasing the bias voltage,
the first step occurs in the current when the electrons tunnel first through more strongly coupled level.
However, with further increase of the bias voltage,
the weakly coupled level becomes occupied and the current drops,
giving rise to corresponding negative differential conductance, see \fig{Fig:5}(a).

When the Hubbard parameter $U$ is nonzero, the electronic structure of the dot is
strongly modified, as shown in \fig{Fig:2}(b), and there are no zero-energy states.
Instead, there are four almost degenerate levels of a finite positive (and negative) energy.
Accordingly, the conductance spectrum shown in \fig{Fig:6}(a) is now greatly changed.
The central diamond (corresponding to the charge neutral dot)
is much larger, as the gate shift for the dot to be charged with one excess
electron involves both the charging energy $E_C$ and the energy of the discrete level.
Further diamonds are equal since they correspond to populating
next levels from the set of four degenerate levels,
and therefore their size is determined by $E_C$ only.
The corresponding Fano factor is presented in \fig{Fig:6}(b).
As before, the shot noise is generally super-Poissonian (or close to Poissonian)
in the Coulomb blockade regions and sub-Poissonian outside these regions.
One should also note that now the largest Fano factor is observed in the central
diamond around $\e = 0$, which is then very close to (or slightly above) unity,
while in the case of $U=0$ super-Poissonian shot noise was mainly present
in two largest Coulomb diamonds.

By comparing \fig{Fig:5} and \fig{Fig:6}, one can conclude that
the key impact of the edge magnetism on the conductance spectra
is the appearance of energy gap and the associated enlarged central diamond (blockade region).
Moreover, if there are no magnetic moments ($U=0$),
the conductance spectrum is generally more complex than that for $U=t$.
It is because the inter-level spacings in the vicinity of the charge neutrality point
are then much smaller than those for $U=t$ (see \fig{Fig:2}) and,
consequently, the depicted spectra are more poorly resolved.
Moreover, there are then also small regions where the differential conductance becomes negative,
which follows from different couplings of particular discrete levels to the electrodes.

\subsubsection{The role of spin-orbit interaction}

\begin{figure}[t!]
  \includegraphics[width=1\columnwidth]{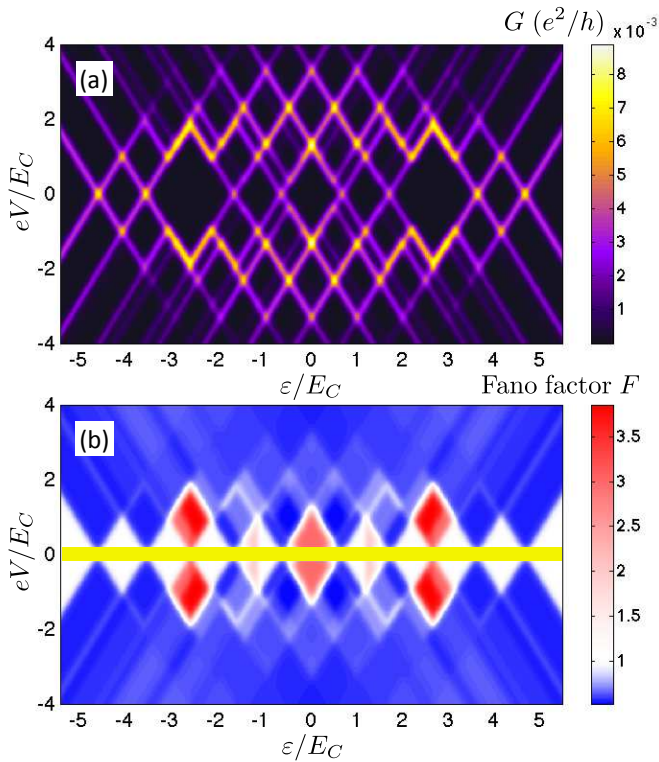}
  \caption{\label{Fig:7}
  (Color online) The bias and level position dependence
  of the differential conductance (a) and the Fano factor (b)
  calculated for $U=0$ and $\tSO/t=0.025$.
  The other parameters are the same as in \fig{Fig:5}.}
\end{figure}

When the spin-orbit interaction is included,
the electronic structure of the dot becomes modified.
First, for $U=0$ there are then no zero energy states.
As shown in \fig{Fig:3}(b), the spin-orbit interaction
lifts partially the degeneracy and two states are shifted up in energy,
while the other two states are shifted down.
Accordingly, the central diamond in the conductance spectrum
[see \fig{Fig:7}(a)] is now larger than the neighboring ones,
contrary to the corresponding situation with no spin-orbit interaction
[see \fig{Fig:5}(a)], where the central blockade diamond is of the same size as the adjacent ones.

\begin{figure}[t!]
  \includegraphics[width=1\columnwidth]{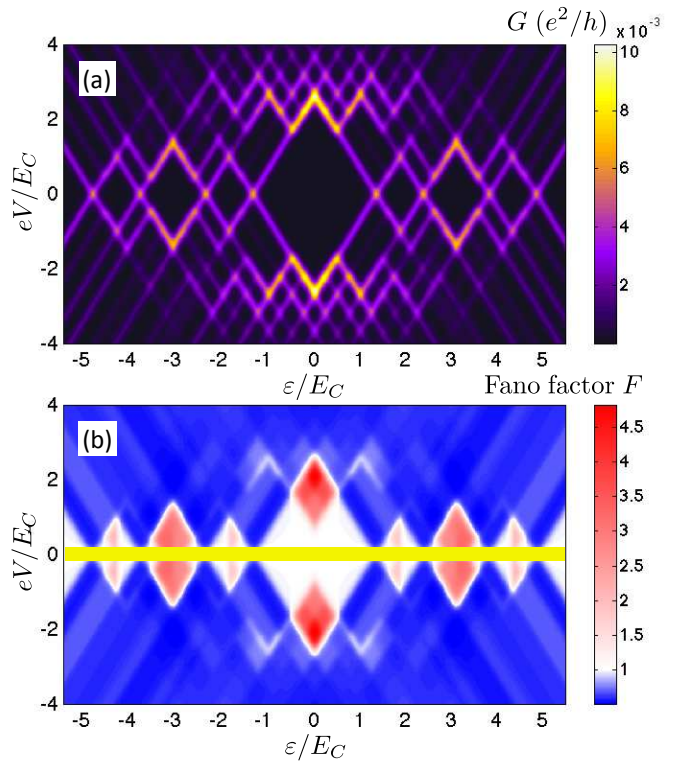}
  \caption{\label{Fig:8}
  (Color online) The same as in \fig{Fig:7}
  calculated for magnetic GLQD ($U=t$) and for $\tSO/t=0.025$.}
\end{figure}

When $U$ is nonzero, the electronic structure is qualitatively similar to that for $U=0$, except that the central gap
in the spectrum (the gap between the first pairs of states with positive and negative energy)
becomes increased, see \fig{Fig:3}(b).
This significantly changes the conductance spectra, as can be seen in \fig{Fig:8}.
Consequently, the central Coulomb blockade diamond is large,
while the next one is determined only by
the charging energy $E_C$, and is therefore much smaller.
The third Coulomb blockade diamond is only slightly larger
than the second one, as it is determined by the charging energy
and a small shift of the third energy level relative to the first two ones.
As before, the spectrum is symmetric with respect to the particle-hole symmetry point,
corresponding to $\varepsilon = 0$. Interestingly, it turns out that the complexity of the spectra gets
enhanced with the increasing $\tSO$ parameter.
The reason is that magnetic moments get then reduced,
and the energy level separations become smaller and smaller.

The corresponding Fano factor is shown in
\fig{Fig:7}(b) in the case of $U=0$ and in \fig{Fig:8}(b) for $U=t$.
As before, the shot noise is generally Poissonian or super-Poissonian in the
blockade regions and sub-Poissonian out of these regions.
However, a closer look reveals certain differences
as compared to the case of $\tSO = 0$,
which are most pronounced in the Coulomb blockade regime.
In the case of finite spin-orbit interaction, the Fano factor
takes the largest values in the central Coulomb diamond,
irrespective of the value of $U$. This is associated with the fact that
there is an energy gap in the spectrum around zero energy
for both finite and zero $U$, cf. \fig{Fig:3}(b).
For finite $\tSO$, super-Poissonian shot noise can be also found
in other Coulomb diamonds, see Figs. \ref{Fig:7}(b) and \ref{Fig:8}(b).
Moreover, out of the Coulomb blockade regime,
the Fano factor is now always much below the Poissonian value, which is
contrary to the case of $\tSO=0$, for which
the shot noise is larger, though still $F\lesssim 1$.

\section{Concluding remarks}\label{Sec:sum}

In this paper we have addressed the problem of magnetic and
transport properties of quantum dots made of graphene-like materials.
First, by using exact diagonalization, the intrinsic spin-orbit interaction was
shown to have a strong impact on the edge magnetic moments of the dots.
In the ground state configuration,
these edge magnetic moments in the presence of intrinsic spin-orbit interaction
are oriented in the systems's plane, while states with perpendicular
moments correspond to a higher energy.
Second, both spin-orbit and on-site Coulomb interactions were found to contribute to the overall
energy gap at the Fermi level of an isolated dot.
Third, the modification of the electronic spectrum by these interactions
was shown to have significant impact on the transport properties
of the graphene-like dot.

By employing the real-time diagrammatic technique,
we have also studied the behavior of the differential conductance
and the Fano factor as a function of bias and gate voltages.
The graphene-like quantum dot was assumed
to be weakly coupled to external leads and
transport was calculated by including the sequential tunneling processes.
The stability diagrams and the Fano factor spectra
have been shown to include some
information on the edge magnetism (on-site Coulomb interaction)
and on the intrinsic spin-orbit coupling.
More specifically, both these interactions remove
the degeneracy of zero energy states and therefore contribute
to the increased size of central Coulomb blockade diamonds.
The size of the diamond is determined by the charging energy
and the energy separation of two successive relevant energy levels.
However, if some diamonds are equal in size,
the corresponding differential conductances might be
different. This is because, in general, different discrete energy levels
are coupled to the leads with different strengths.
In certain cases, this may give rise to negative differential conductance.

In summary, the main result of this study is a detailed analysis
of the combined effect of the intrinsic spin orbit interaction
and the resulting in-plane edge magnetization
on the Coulomb blockade spectra.
We have shown that the Coulomb blockade spectra
give important bits of information on the involved
edge magnetism and its easy plane.
On the one hand, our calculations have shown that protected edge
states do not appear in the case of small graphene-like
flakes with the in-plane edge magnetization.
This is in accordance with the energy band structure
results reported for infinitely long ribbons, \cite{Lado2014}
which clearly show that no crossing energy levels appear
in the band gap in that case (in contrast to the out-of-plane case).
On the other hand, we have demonstrated that the in-plane magnetization,
as well as the corresponding energy gap is relatively
robust against the intrinsic spin orbit interaction.
From the perspective of potential spintronic applications
the first fact is rather disappointing, but the second one is positive.

\begin{figure}[t!]
  \includegraphics[width=0.9\columnwidth]{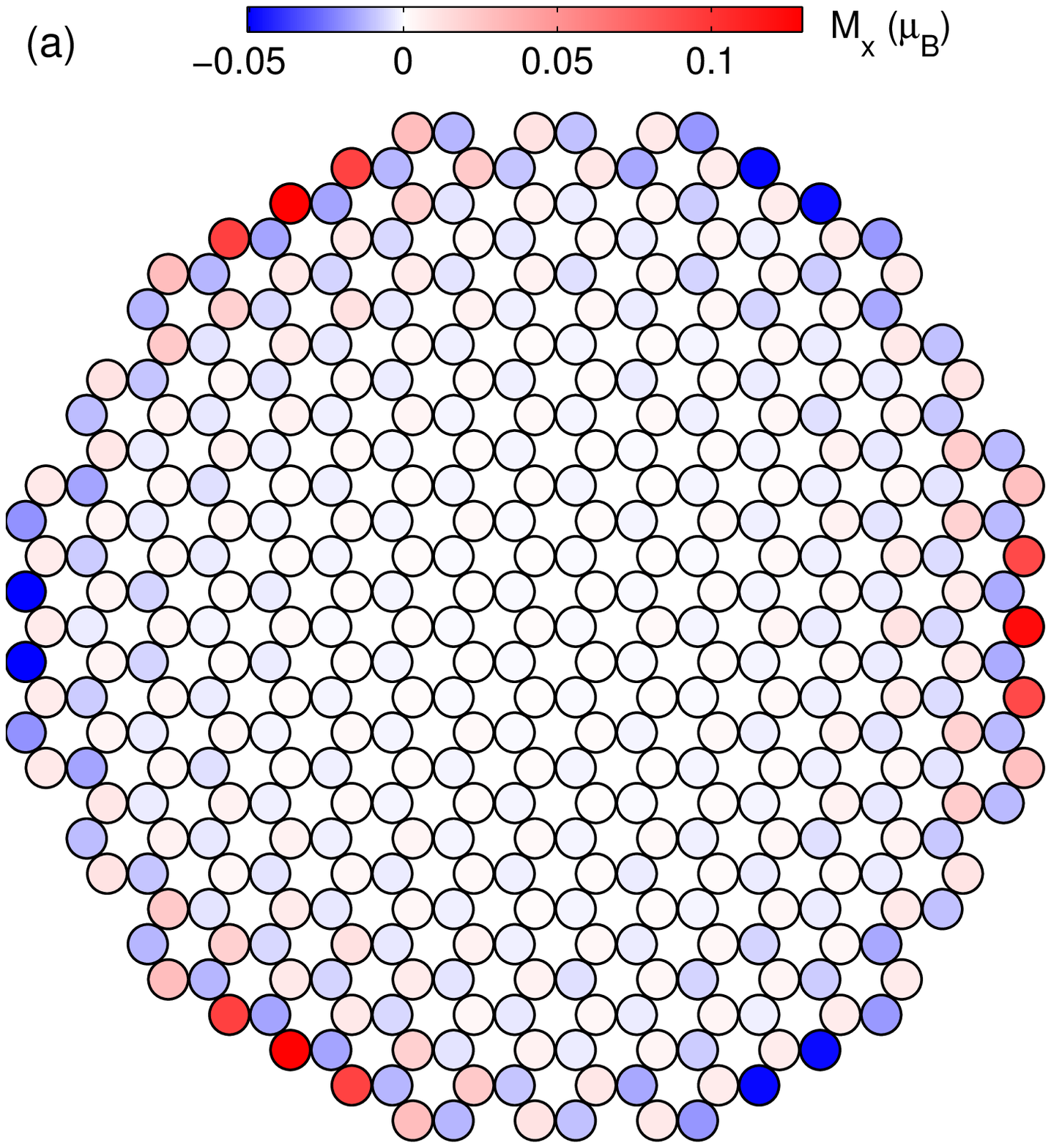}
  \includegraphics[width=0.8\columnwidth]{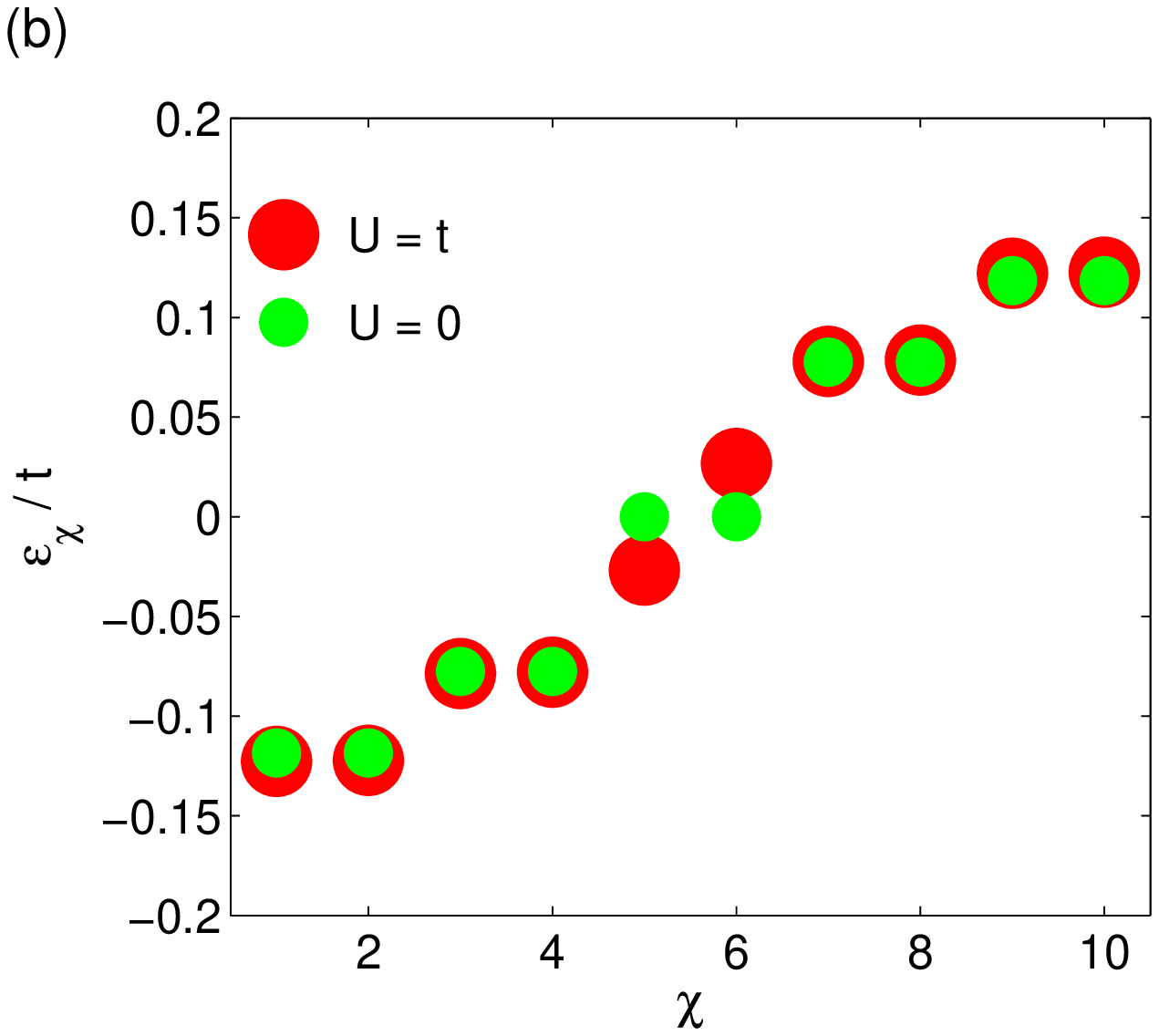}
  \caption{\label{Fig:9}
  (Color online)
  Distribution of magnetic moments on individual atoms (a),
  and the eigenvalues (b) around the charge neutrality point
  calculated for the case of $U=t$ and $\tSO / t = 0.025$
  for graphene-like nanodisk of irregular shape.
  For comparison, in panel (b) we also show the eigenvalues calculated for $U=0$.}
\end{figure}

\begin{acknowledgments}
This project was supported by the Polish National Science Centre
from funds awarded through the decision No. DEC-2013/10/M/ST3/00488.
\end{acknowledgments}

\appendix*
\section{Effect of edge irregularities\label{Appendix}}

It is important to test whether the findings presented in the main text are
still valid for other less simple geometries than those of rectangular
flakes. To do this we consider a graphene-like flake of a nanodisk (circular) shape,
large enough to contain both zigzag and armchair type arrangements of edge atoms,
see \fig{Fig:9}(a).
The corresponding spin and energy structure
calculated for $\tSO/t$=0.025 and $U=t$
is presented in Figs. \ref{Fig:9}(a) and (b), respectively.
It can be clearly seen that indeed magnetic moments
are located at zigzag-type fragments of the circumference.
Moreover, the magnetic moments calculated for other $\tSO$ values
show that the edge magnetic moments
decrease with increasing $\tSO$
and the in-plane magnetic configuration is always energetically
more favorable than the out-of-plane one.
Similarly to the case of rectangular flakes, also the nanodisk's edge magnetic moments
of the in-plane configuration are greater than
those corresponding to the (unstable) out-of-plane configuration.

All this indicates that the effects described
in the main part of the paper are relatively
resistant to structural irregularities and shape of the flake's boundary.
The only difference between the results in the case of rectangular and circular flakes
is that the latter magnetic configuration is ferrimagnetic rather than
antiferromagnetic, with magnetic moments reduced by a few tens of percent.
This difference results from the increased imbalance of the A and B sublattices,
as well as from short lengths of the zigzag-type edge fragments in the nanodisk case.

As concerns the Coulomb blockade stability diagrams,
they strictly reflect the underlying energy spectra.
Similarly as it was done in the case of rectangular flakes
discussed in the main text,
from the size of the Coulomb diamonds 
one can again extract some information
about the intrinsic parameters of the flake.

\end{document}